\documentclass[twocolumn,twoside,slac_two]{revtex4}
\usepackage{graphicx}
\usepackage{fancyhdr}
\def\Bbar    {\kern 0.18em\overline{\kern -0.18em B}{}}
\def\Kbar    {\kern 0.18em\overline{\kern -0.18em K}{}}
\def\babar{\mbox{\sl B\hspace{-0.4em} {\small\sl A}\hspace{-0.37em} \sl B\hspace{-0.4em} {\small\sl A\hspace{-0.02em}R}}}
\def\babarcap {\mbox{\sl B\hspace{-0.4em} {\footnotesize\sl A}\hspace{-0.37em} \sl B\hspace{-0.4em} {\footnotesize\sl A\hspace{-0.02em}R}}}
\def\babarabs {\mbox{\sl B\hspace{-0.4em} {\scriptsize\sl A}\hspace{-0.37em} \sl B\hspace{-0.4em} {\scriptsize\sl A\hspace{-0.02em}R}}}

\begin{document}

\title{Rare Leptonic ${\mathbf B}$ and ${\mathbf b \to s \ell^+ \ell^-}$ 
Decays at ${\mathbf B}$-factories}

%

\author{Jack L. Ritchie \enskip (from the \babar\ Collaboration)}
\affiliation{University of Texas at Austin, Austin, Texas, USA}

\begin{abstract}
We review recent results from the \babarabs\ and Belle experiments on
rare electroweak $B$-meson decays, with emphasis on those occuring through
flavor changing neutral current interactions of the type $b \to s \ell \bar\ell$.
The recent results
include measurements of the isospin asymmetry and lepton forward-backward asymmetry
in $B \to K^* \ell^+ \ell^-$ decays from \babarabs, a search for $B \to \pi \ell^+ \ell^-$
from Belle, and searches for $B \to K^{(*)} \nu \bar\nu$ from \babarabs.  We also
briefly review the status of $B^0 \to \ell^+ \ell^-$ and $B^+ \to \ell^+ \nu$
searches.
\end{abstract}

\maketitle

\thispagestyle{fancy}


\section{Introduction}
The study of flavor-changing neutral current (FCNC) decays has been a fruitful
avenue of research throughout the history of particle physics.  Today
rare $B$-meson decays based on the process $b \to s \ell^+ \ell^-$ (where $\ell = e$ or $\mu$)
provide one of the most promising probes for new physics.
Both $B$-factory experiments, \babar\ and Belle, have observed decays of this type and
reported measurements which probe their underlying structure.

\begin{figure}[h]
\centering
\includegraphics[width=80mm]{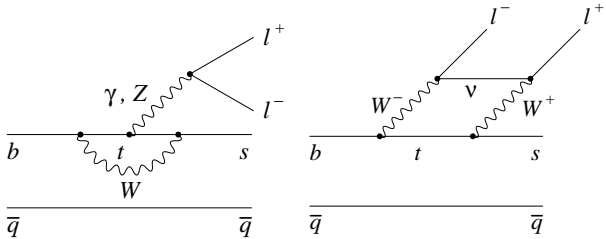}
\caption{The electroweak penguin (left) and $W$-box (right) diagrams
responsible for $B \to K^{(*)}\ell^+ \ell^-$ decays.} 
\label{fig-slldiags}
\end{figure}

The decays proceed through loop diagrams such as those shown in Figure~\ref{fig-slldiags}.
The theoretical treatment of $b \to s \ell^+ \ell^-$ transitions in the Standard Model (SM) invokes
an effective field theory in which the Hamiltonian is a sum of terms,
where each term consists of CKM factors and a Wilson coefficient multiplying an
operator that is formed from the light quark and lepton fields.  The Wilson coefficients,
obtained by integrating out the heavy particles, characterize the
short-distance physics in these decays.    New physics (e.g., SUSY) would
modify the Wilson coefficients by providing new particles inside the loops and may
modify the Hamiltonian by adding scalar or pseudoscalar terms.  
To account for QCD effects that mix the operators, so-called effective Wilson
coefficients are defined, which are functions of a renormalization scale~$\mu$ (typically
taken to be 4.6~GeV in the $\overline{MS}$ scheme)
and the squared dilepton mass $q^2 = m^2_{\ell \ell}$.  
Measurements of $b \to s \ell^+ \ell^-$
decays mainly
probe the effective Wilson coefficients
$\tilde C_7$, $\tilde C_9$, and $\tilde C_{10}$.  Fully inclusive measurements are
not possible, and the background environment is very difficult 
for analyses that combine multiple exclusive modes, so it is conventional to focus
on the exclusive modes $B \to K \ell^+ \ell^-$ and $B \to K^\ast \ell^+ \ell^-$
(denoted together as $B \to K^{(\ast)} \ell^+ \ell^-$).  Form factor
uncertainties complicate the interpretation of exclusive measurements,
but a number of measurable asymmetries, defined below, are 
relatively insensitive to these uncertainties.
An up-to-date summary of the
theoretical status, along with extensive references, can be found in Ref.\cite{Gerald-FPCP}.

The related decays $B \to K^{(\ast)} \nu \bar \nu$ are also of considerable interest, but
experimentally present almost insurmountable problems owing to the missing neutrinos.  Nonetheless,
the $B$-factory experiments have conducted searches, and \babar\ has a recent update.

The decays $B \to \pi \ell^+ \ell^-$ and $B^0 \to \ell^+ \ell^-$ proceed through
essentially the same diagrams, where the $s$ quark is replaced by a $d$ quark,
leading to CKM suppression by a factor $|V_{\rm td}/V_{\rm ts}|^2\simeq 0.04$; in addition,
the two-body decays are helicity suppressed.  Consequently, these decays have quite small
SM branching fractions and have thus far not been observed.  Belle has recently reported a new
limit on $B \to \pi \ell^+ \ell^-$.

The decays $B^+ \to \ell^+ \nu$ proceed through simple $W$-exchange and are
sensitive to the product $f_B |V_{ub}|$.  Combined with a measurement of $|V_{ub}|$ from another source, measurement of any of these branching fractions would provide a good measurement of the
$B$-meson decay constant $f_B$.  For the $\ell = \mu$ and $e$
cases, the branching fractions are quite small due to helicity suppression, although the
$B^+ \to \mu^+ \nu$ decay appears to be only just beyond the reach of the current $B$-factory
experiments.  The status of $B^+ \to \tau^+ \nu$ is presented in another contribution to
this conference \cite{MBarrett}.

Below, the status of these rare FCNC $B$ decays is described, with emphasis on recent results
from Belle and \babar.

\section{$ B \to K^{(*)} \ell^+ \ell^-$ Measurements}

These decays experience interference from the processes
$B \to  K^{(*)} J/\psi  \to   K^{(*)} \ell^+ \ell^-$ and
$B \to  K^{(*)} \psi(2S)  \to   K^{(*)} \ell^+ \ell^-$.  Consequently, it is
necessary to remove events with lepton-pair mass close to the $J/\psi$ or
$\psi(2S)$ peaks.  On the positive side, these processes provide large and
well-understood control samples which can be used to study efficiencies and
to study the characteristics of signal-like events (e.g., to determine PDFs for
fits).  

The main backgrounds to these decays arise from $B$ and $D$ semileptonic decays.
The two leptons can come from the semileptonic decay of both the $B$ and $\overline{B}$
in an event, or from the semileptonic decays of a $B$ and also its daughter $D$.
These backgrounds are suppressed by combining event shape information, vertex information,
and missing energy information into multivariate analysis techniques (such as neural nets).
Another important background, $B \to D \pi$ (followed by $D \to K^{(*)} \pi$) wherein pions
are misidentified as muons, can be explicity vetoed by rejecting events in which one of the
muons, when assigned the pion mass, reconstructs in combination with a kaon 
to the $D$ mass.  Signal can then be
separated from the residual background using maximum likelihood fits, typically utilizing
the differing shapes of signal and background distributions in the quantities 
$\Delta E = E^*_B - E^*_{\rm beam}$ and 
$m_{ES}[{\rm \babar}]=m_{\rm bc}[{\rm Belle}] = \sqrt{E^{*2}_{\rm beam} - p^{*2}_B}$,
where $E^*_B$ and $p^*_B$ are the CM energy and momentum of the reconstructed $B$.

\babar\ has recently reported a series of measurements based on $349 \, {\rm fb^{-1}}$ of
data (384 million $B\overline{B}$ pairs).  These results are described below and are compared
to Belle results where possible.  

The \babar\ analysis of $B \to K^{(*)} \ell^+ \ell^-$ reconstructs 10 submodes,
reflecting five possible $K^{(*)}$ final states ($K^\pm$, $K^0_S$, $K^\pm \pi^\mp$,
$K^\pm \pi^0$, and $K^0_S \pi^\pm$), each paired with $e^+e^-$ or $\mu^+ \mu^-$.
Electrons (muons) are required to have $p>0.3 (0.7) \, {\rm GeV/c}$.  Photons consistent
with bremsstrahlung are combined with the associated electron.  Two $q^2$-bins are defined:
a low-$q^2$ bin 
$0.1 < q^2 < 7.02 \, {\rm GeV^2}$ and a
high-$q^2$ bin $q^2 > 10.24 \, {\rm GeV^2}$ but excluding $12.96 < q^2 < 14.06$ for the $\psi(2S)$.
More $q^2$ bins are desirable, but are not possible with the current event sample.
The so-called pole region ($q^2 < 0.1 \, {\rm GeV^2}$) is excluded due to the $1/q^2$ term in
$B \to K^* \gamma$.

\subsection{Branching Fractions}

The values of the recent \babar\ branching fraction
measurements, which combine both $q^2$ bins
and lepton-pair cases, 
and includes correction for the excluded $J/\psi$ and $\psi(2S)$ regions, are:

\begin{center}
${\cal B}(B \to K \ell^+ \ell^-) = (3.9 \pm 0.07 \pm 0.2) \times 10^{-7},$

\qquad \\

${\cal B}(B \to K^* \ell^+ \ell^-) = (11.1^{+1.9}_{-1.8} \pm 0.7) \times 10^{-7}.$
\end{center}

\begin{figure}[h]
\centering
\includegraphics[width=80mm]{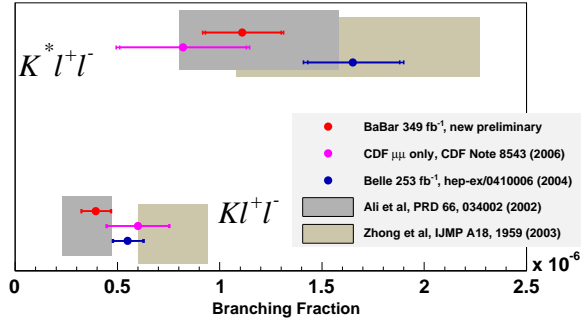}
\caption{Summary of branching fraction results from different
experiments for $B \to K^{(*)}\ell^+ \ell^-$.} 
\label{fig-sllbfs}
\end{figure}

Figure~\ref{fig-sllbfs} shows these results, along with prior measurements from other experiments
and two theoretical results based on the SM.  The experimental results are consistent
with the SM predictions.  

\subsection{Decay Asymmetries}

By reconstructing 10 separate submodes for $B \to K^{(*)}\ell^+ \ell^-$, it becomes
possible to construct a variety of decay asymmetries which test different aspects of
the process.  Direct CP violation, which is expected to be very small in the SM, can be tested by comparing the decay rates for $B$
to $\Bbar$:

$$A_{CP} \equiv { {\cal B}(\Bbar \to \Kbar^{(\ast)} \ell^+ \ell^-) - 
{{\cal B}(B \to K^{(\ast)} \ell^+ \ell^-)} 
\over 
{\cal B}(\Bbar \to \Kbar^{(\ast)} \ell^+ \ell^-) + 
{{\cal B}(B \to K^{(\ast)} \ell^+ \ell^-)}  }
$$.

The recent \babar\ analysis reports:
\begin{center}
$A_{CP}^K =  -0.18^{+0.18}_{-0.18} \pm 0.01,$

\qquad \\

$A_{CP}^{K*} =  0.01^{+0.16}_{-0.15} \pm 0.01$.
\end{center}

A lepton-flavor asymmetry (test of $\mu$-$e$ universality) can be defined by the ratio:
$$
R \equiv {{\cal B}(B \to K^{(*)} \mu^+ \mu^-) \over {\cal B}(B \to K^{(*)} e^+ e^-)}.
$$

Models that would enhance $B_s \to \mu^+ \mu^-$, such as SUSY with a Higgs at large $\tan \beta$,
would also enhance $R$ somewhat.  At the current level of statistics, the test is not very
restrictive, but is consistent with the SM expectation of unity.  The recent \babar\ results
are (for $q^2 > 0.1 \, {\rm GeV^2}$):
\begin{center}
$R_K =  0.96^{+0.44}_{-0.34} \pm 0.05,$

\qquad \\

$R_{K*} =  1.37^{+0.53}_{-0.40} \pm 0.09$.
\end{center}

\begin{figure}[h]
\centering
\includegraphics[width=80mm]{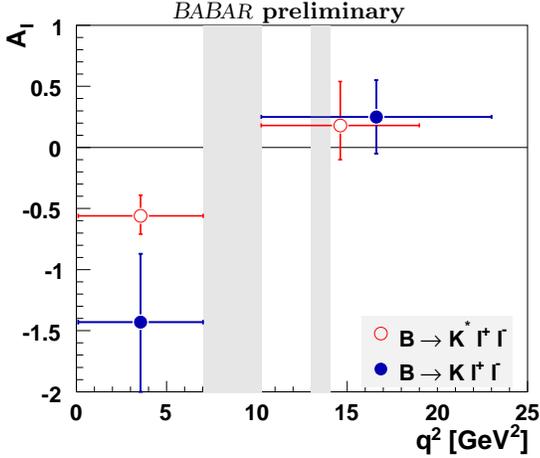}
\setlength{\unitlength}{1in}
\begin{picture}(0,0) \put(-0.6,2.55){\textbf{\babarcap\ preliminary}} \end{picture}
\caption{The isospin asymmetry  versus $q^2$ from
\babarcap: $B \to K \ell^+ \ell^-$ (solid dots, blue);
$B \to K^* \ell^+ \ell^-$ (open circles, red).} 
\label{fig-sllai}
\end{figure}

The most interesting and potentially important recent
results address the isospin asymmetry,
which compares the decay of neutral to charged $B$'s:

$$A_{I} \equiv { {\cal B}(B^0 \to K^{(\ast)0} \ell^+ \ell^-) - 
r{{\cal B}(B^\pm \to K^{(\ast)\pm} \ell^+ \ell^-)} 
\over 
{\cal B}(B^0 \to K^{(\ast)0} \ell^+ \ell^-) + 
r{{\cal B}(B^\pm \to K^{(\ast)\pm} \ell^+ \ell^-)}  },
$$
where $r = \tau_0/\tau_+$ is the ratio of $B^0$ to $B^+$ lifetime.
The value of $A_I$ is expected to be close to zero in the SM, although
at low-$q^2$ some deviation is expected (up to about 10\%);  in particular,
the sign of this deviation depends on the sign of $\tilde C_7$ \cite{FM}.
Figure~\ref{fig-sllai} shows the $A_I$ result in the two $q^2$-bins in the
recent \babar\ analysis.  A significant deviation from zero is observed in the low-$q^2$ bin
for both $K$ and $K^*$ ($\sim 3\sigma$ in each case).

\begin{figure}[ht]
\centering
\includegraphics[width=80mm]{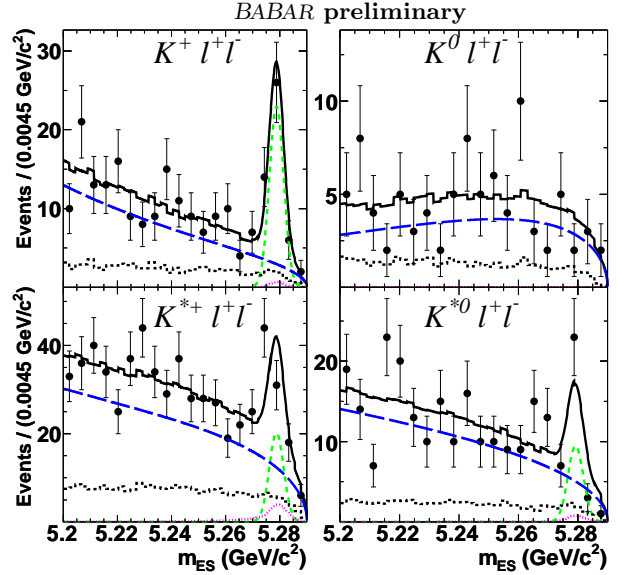}
\setlength{\unitlength}{1in}
\begin{picture}(0,0) \put(-0.4,3.05){\textbf{\babarcap\ preliminary}} \end{picture}
\caption{$m_{ES}$ distributions for the low-$q^2$ bin: $K^\pm \ell^+ \ell^-$ (upper left);
$K^0 \ell^+ \ell^-$ (upper right);
$K^{*\pm} \ell^+ \ell^-$ (lower left);
$K^{*0} \ell^+ \ell^-$ (lower right).
Fit results are superimposed: full fit (solid blue), signal (black dashed),
combinatorial background (magenta dashed), misidentified muons (green dotted), 
peaking backgrounds (red dotted).} 
\label{fig-aifits}
\end{figure}

The underlying fits to $m_{ES}$ distributions for the low-$q^2$ bin 
are shown in Figure~\ref{fig-aifits}.  The corresponding partial
branching fractions are (for $0.1 < q^2 < 7.02 \, {\rm GeV^2}$):

\begin{center}
${\cal B}(B^\pm \to K^\pm \ell^+ \ell^-) = (2.5 \pm 0.5 \pm 0.1) \times 10^{-7},$

\qquad \\

${\cal B}(B^0 \to K^0 \ell^+ \ell^-) < 0.9 \times 10^{-7}$ (90\% CL),

\qquad \\

${\cal B}(B^\pm \to K^{*\pm} \ell^+ \ell^-) = (9.8^{+2.6}_{-2.4} \pm 0.6) \times 10^{-7},$

\qquad \\

${\cal B}(B^0 \to K^{*0} \ell^+ \ell^-) = (2.6^{+1.1}_{-1.0} \pm 0.2) \times 10^{-7}.$

\end{center}

Subsequent to HQL08, Belle presented $A_I$ measurements based on a $625 \, {\rm fb^{-1}}$
dataset at ICHEP08 \cite{belle-ichep08}.   Belle's new
results also indicate negative $A_I$ values for $q^2$ below the $J/\psi$, but Belle observes
less pronounced deviations from unity than \babar.
Negative $A_I$ values at low $q^2$ tends
to favor a flipped-sign for $\tilde C_7$.

\subsection{Angular Analysis}

Angular distributions as functions of $q^2$,
particularly the forward-backward lepton asymmetry $A_{FB}$, are particularly sensitive 
to possible
new physics.  This asymmetry, defined as follows, basically measures the tendency of
the $\ell^+ (\ell^-)$ to be in the same hemisphere as the $\Bbar(B)$ when viewed from the dilepton rest frame:
$$
A_{FB}(q^2) = { 1 \over {d \Gamma  \over dq^2} } 
\int^1_{-1} d \cos \theta_l {d^2 \Gamma  \over dq^2 d \cos \theta_l} {\rm sgn}(\cos \theta_l),
$$
where $\Gamma$ is the $B \to K^* \ell^+ \ell^-$ decay width and $\theta_\ell$ is the angle between
the $\ell^-$ and $B$ in the $\ell^+ \ell^-$ rest frame.  

An additional angular variable of importance is the fraction of longitudinal $K^*$ polarization,
$F_L$.  The quantity $F_L$ has some sensitivity to new physics and also affects the angular
distributions of $\theta_l$ which must be fit to determine $A_{FB}$.
Extraction of $F_L$ and $A_{FB}$ from $B \to K^* \ell^+ \ell^-$ candidate 
events requires an understanding
of the angular correlations of background events.  These can be studied using
$B \to K^* \mu^\pm e^\mp$ events in data, as well as Monte Carlo simulations.
The analysis procedure and fitting method can be validated using the
$B \to J/\psi [\psi(2s)] K^*$ control samples discussed earlier.
In addition, a null result for $A_{FB}$ is expected in $B \to K \ell^+ \ell^-$, providing
another check.

\begin{figure}[t]
\centering
\includegraphics[width=80mm]{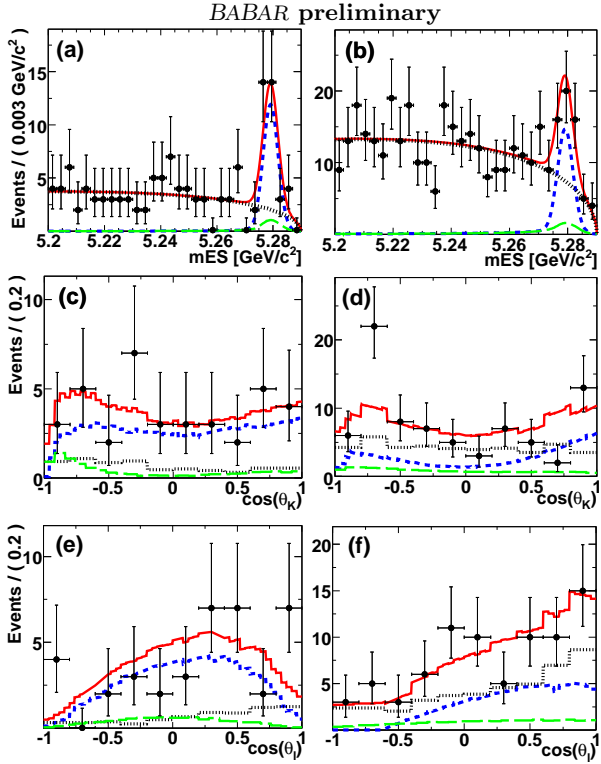}
\setlength{\unitlength}{1in}
\begin{picture}(0,0) \put(-0.5,4.05){\textbf{\babarcap\ preliminary}} \end{picture}
\caption{\babarcap\ fit results for determining $F_L$ and $A_{FB}$ in two $q^2$ bins.
(a) and (b) $m_{ES}$ distributions in low- and high-$q^2$ bins, respectively, used
to fit signal and background yields;
(c) and (d) $\cos \theta_K$ distributions fit to determine $F_L$;
(e) and (f) $\cos \theta_l$ distributions fit to determine $A_{FB}$.
The total fit is shown in red (solid); signal in blue (dashed);
combinitorial background in black (dots); and peaking backgrounds in green (long dashes).
} 
\label{fig-afbfits}
\end{figure}

\babar\ performs a three-step procedure, the results of which are shown in Figure~\ref{fig-afbfits}.
The first fit determines signal and background yields; the second determines $F_L$; and the
third determines $A_{FB}$.  The results, in the low- and high-$q^2$ bins, are:

\begin{center}
$F_L^{\rm low} =  0.35 \pm 0.16 \pm 0.04,$

\qquad \\

$F_L^{\rm high} =  0.71^{+0.20}_{-0.22} \pm 0.04,$

\qquad \\

$A_{FB}^{\rm low} =  0.24^{+0.18}_{-0.23} \pm 0.05,$

\qquad \\

$A_{FB}^{\rm high} =  0.76^{+0.52}_{-0.32} \pm 0.07$.
\end{center}

\begin{figure}[htb]
\centering
\includegraphics[width=80mm]{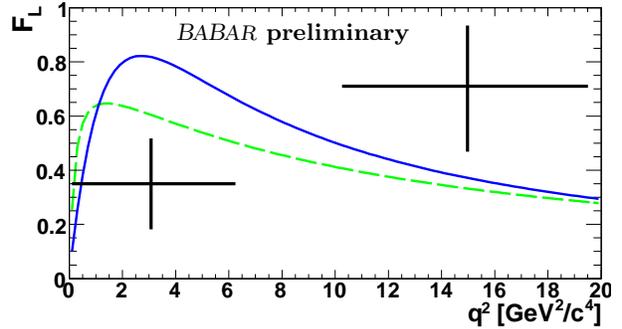}
\setlength{\unitlength}{1in}
\begin{picture}(0,0) \put(-0.7,1.65){\textbf{\babarcap\ preliminary}} \end{picture}
\caption{\babarcap\ $F_L$ results in two $q^2$ bins.  The solid (blue) line is the SM
expectation.   The dashed green line shows a flipped-sign $\tilde C_7$ model.
} 
\label{fig-fl}
\end{figure}

Obviously these results are statistically limited, even in two $q^2$ bins.
Nonetheless, they can be compared
with SM expectations.  Figure~\ref{fig-fl} shows the $F_L$ measurements along with the
SM expectation, as well as a flipped-$\tilde C_7$ scenario.

\begin{figure}[htb]
\centering
\includegraphics[width=80mm]{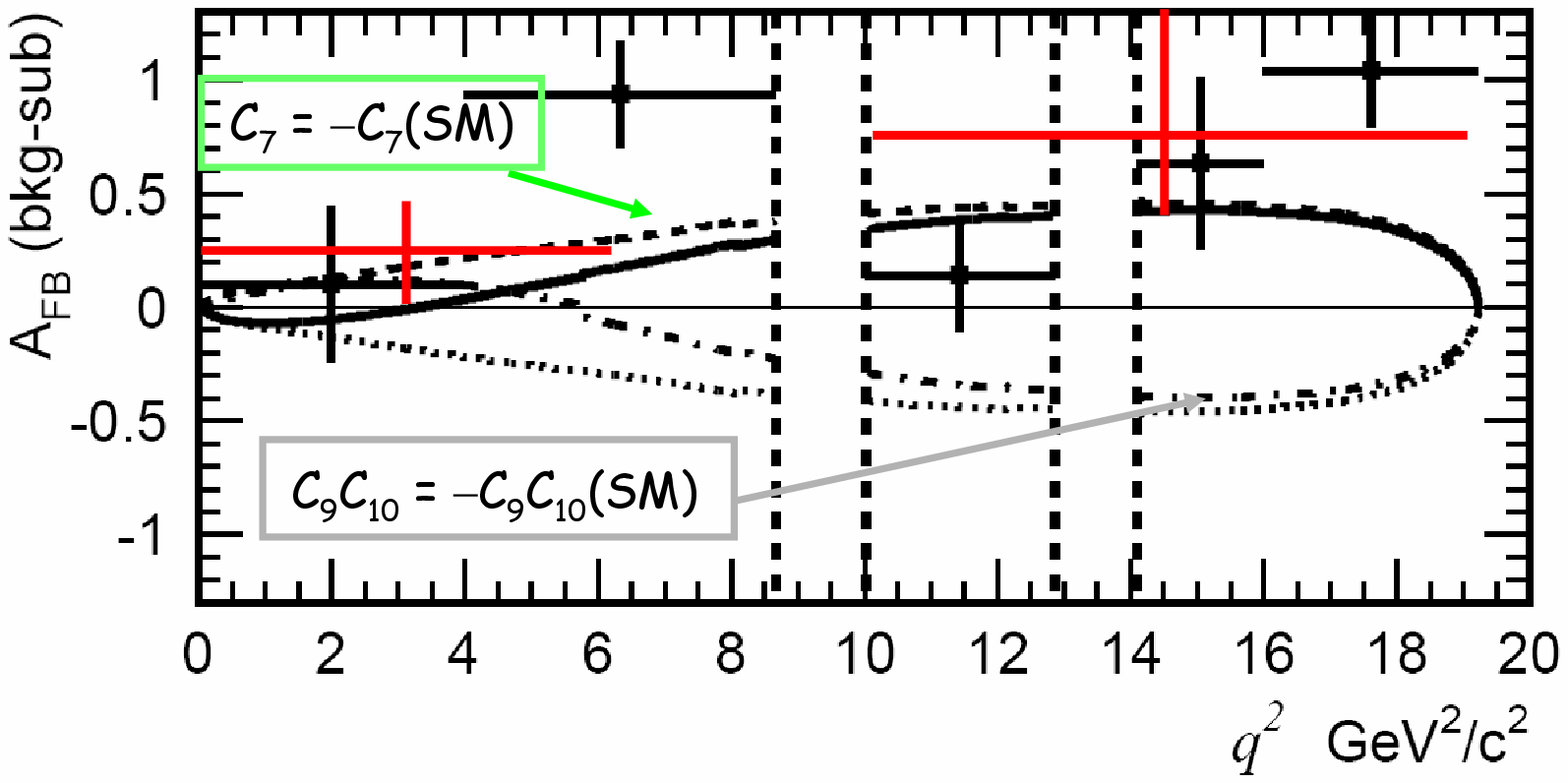}
\caption{\babarcap\ Preliminary \babarcap\ $A_{FB}$ results in two $q^2$ bins in red.  
The black points are from Belle \cite{belle-afb}.  The solid black curve is the SM
expectation.   The dashed curve shows a flipped-sign $\tilde C_7$ model.
The dotted curve and the dot-dashed curve show models with the sign of $\tilde C_9 \tilde C_{10}$
flipped.
} 
\label{fig-afb}
\end{figure}

The $A_{FB}$ results are shown in Figure~\ref{fig-afb}, which overlays the recent
\babar\ results on prior Belle results \cite{belle-afb}.  The \babar\ and Belle
results are consistent, and tend to favor positive values of $A_{FB}$, particularly
at large $q^2$.  This disfavors the models, shown in Figure~\ref{fig-afb}, with the
sign of $\tilde C_9 \tilde C_{10}$
flipped.  Subsequent to HQL08, Belle presented updated $A_{FB}$ results \cite{belle-ichep08}
based on $625 \, {\rm fb^{-1}}$ at ICHEP08;  the updated results are consistent
with the prior Belle results and the recent \babar\ results.

\section{Searches for $B \to \pi \ell^+ \ell^-$}

Belle recently reported a new search for $B \to \pi \ell^+ \ell^-$
based on 657~million $B \Bbar$ pairs (about $607 \, {\rm fb^{-1}}$) \cite{belle-pill}.
Continuum and semileptonic $B$ decay backgrounds are suppressed using likelihood ratios
that combine event shape, vertex, and other information.  Two-dimensional maximum likelihood
fits are performed in the variables $\Delta E$ and $m_{\rm bc}$.  No signals are observed and
90\% confidence level limits are set on $B^+ \to \pi^+ \ell^+ \ell^-$ and
$B^0 \to \pi^0 \ell^+ \ell^-$, as well as an isospin-averaged combination of the two.
These new limits are listed in Table~\ref{table-pill}, along with previous limits
from \babar\ \cite{babar-pill}.
The table also includes SM expectations \cite{theory-pill}.
It is noteworthy that the Belle upper limit on $B^+ \to \pi^+ \ell^+ \ell^-$ is approaching
the expected SM branching fraction. 

\begin{table}[h]
\begin{center}
\caption{Upper limits (90\% CL) on ${\cal B}(B \to \pi \ell^+ \ell^-)$ from Belle and \babarcap,
along with SM expectations.}
\begin{tabular}{|l|c|c|c|}
\hline \textbf{Mode} & \textbf{SM} & \textbf{Belle} &
\textbf{\babarcap}
\\
\hline 
$B^+ \to \pi^+ \ell^+ \ell^-$ & $3.3\times 10^{-8}$ & $<4.9\times 10^{-8}$ & $<12\times 10^{-8}$ \\
\hline 
$B^0 \to \pi^0 \ell^+ \ell^-$ & $1.7\times 10^{-8}$ & $<15.4\times 10^{-8}$ & $<12\times 10^{-8}$ \\
\hline 
$B \to \pi \ell^+ \ell^-$ & $3.3\times 10^{-8}$ & $<6.2\times 10^{-8}$ & $<9.1\times 10^{-8}$ \\
\hline
\end{tabular}
\label{table-pill}
\end{center}
\end{table}

\section{Searches for $B \to K^{(*)} \nu \bar \nu$}

With missing neutrinos, $B \to K^{(*)} \nu \bar \nu$ decays are particularly difficult
to isolate, and backgrounds are daunting.  To clean up events, both \babar\ and Belle
have performed these searches by requiring one $B$ in the event to be fully reconstructed,
thereby removing continuum events and making it possible to assign particles in the event
either to the tag $B$ or the signal candidate.  \babar\ has reported recent results 
for both $B \to K \nu \bar \nu$ (based on $319 \, {\rm fb^{-1}}$) and
$B \to K^* \nu \bar \nu$ (based on $413 \, {\rm fb^{-1}}$) in separate analyses, both of
which rely on reconstructing one $B$ via $B \to D^{(*)} \ell \nu$.
The $B \to K \nu \bar \nu$ search utilizes a Random Forest algorithm \cite{RandForest}
to separate signal
and background, while the $B \to K^* \nu \bar \nu$ analysis utilizes a maximum likelihood
fit that relies on the differences
in distribution of extra energy
in the event between signal and background.  The results from these searches are listed in Table~\ref{table-knn},
along with the SM expectations \cite{theory-knn} and the current Belle limits \cite{belle-knn}.

\begin{table}[h]
\begin{center}
\caption{Upper limits (90\% CL) on ${\cal B}(B \to K^{(*)} \nu \bar \nu)$ from Belle and \babarcap,
along with SM expectations.  The \babarcap\ results are preliminary.}
\begin{tabular}{|l|c|c|c|}
\hline \textbf{Mode} & \textbf{SM} & \textbf{Belle} &
\textbf{\babarcap}
\\
\hline 
$B^+ \to K^{*+} \nu \bar \nu$ & $1.3\times 10^{-5}$ & $<14\times 10^{-5}$ & $<9\times 10^{-5}$ \\
\hline 
$B^0 \to K^{*0} \nu \bar \nu$ & $1.3\times 10^{-5}$ & $<34\times 10^{-5}$ & $<21\times 10^{-5}$ \\
\hline 
$B^+ \to K^+ \nu \bar \nu$ & $0.4\times 10^{-5}$ & $<1.4\times 10^{-5}$ & $<4.2\times 10^{-5}$ \\
\hline
$B^0 \to K^0 \nu \bar \nu$ & $0.4\times 10^{-5}$ & $<16\times 10^{-5}$ &  \\
\hline
\end{tabular}
\label{table-knn}
\end{center}
\end{table}

\section{Searches for $B^0 \to \ell^+ \ell^-$}

While these decays are sensitive to the same electroweak penguin and $W$-box diagrams as the
other decays discussed above, the prospects of observing them at the SM level are rather
remote.  The
decay $B^0 \to \tau^+ \tau^-$
is experimentally very challenging due to the difficulties
associated with multiple missing neutrinos.  And while the experimental signatures of
$B^0 \to \mu^+ \mu^-$ and $B^0 \to e^+ e^-$ are nearly ideal, these decays are helicity suppressed
to levels that make them inaccessible for the $B$-factories.  The SM branching fractions
are expected to be about $10^{-7}$ for $\tau^+ \tau^-$,
$10^{-10}$ for $\mu^+ \mu^-$,
and $10^{-15}$ for $e^+ e^-$.
Non-SM scalar interactions would not be helicity suppressed, so that there is a window
for new physics above the SM level.  Thus, it is worthwhile to search for these modes even though
the SM level remains inaccessible.
The most promising opportunity,
particularly for the $\mu^+ \mu^-$ mode (including 
$B_s \to \mu^+ \mu^-$), occurs at hadron colliders where the large hadronic $B$-production
cross section provides a major advantage.
Table~\ref{table-ll} summarizes the current status of these modes.

\begin{table}[h]
\begin{center}
\caption{Upper limits (90\% CL) on ${\cal B}(B^0 \to \ell^+ \ell^- )$ from Belle, \babarcap,
and CDF.}
\begin{tabular}{|l|c|c|c|}
\hline \textbf{Mode} & \textbf{Belle} & \textbf{\babarcap} &
\textbf{CDF}
\\
\hline 
$\tau^+ \tau^-$ &              & $<4.1\times 10^{-3}$\cite{babar-tautau} &  \\
\hline 
$\mu^+ \mu^-$ & $<1.6\times 10^{-7}$\cite{belle-ll} & $<5.2\times 10^{-8}$\cite{babar-ll} & $<1.8\times 10^{-8}$\cite{cdf-mumu} \\
\hline 
$e^+ e^-$ & $<1.9\times 10^{-7}$\cite{belle-ll} & $<1.1\times 10^{-7}$\cite{babar-ll} & \\
\hline
\end{tabular}
\label{table-ll}
\end{center}
\end{table}

\section{Searches for $B^+ \to \ell^+ \nu$}

$B^+ \to \ell^+ \nu$ decays would occur via the tree-level exchange of a $W$-boson.
The branching fractions are precisely predictable in the SM
given the relevant CKM factor, $|V_{ub}|$, and the $B$-meson decay constant $f_B$.  
$|V_{ub}|$ is best measured in $b \rightarrow u \ell \nu$ semileptonic decays;  a review
of the latest measurements can be found in this proceeding \cite{Petrella}. 

Considerable progress has been made by both Belle and \babar\ toward measurements
of $B^+ \to \tau^+ \nu$.  This is the topic of another contribution to this proceeding\cite{MBarrett}
and will not be discussed here.  Table~\ref{table-lnu} summarizes the status of the other modes.
It is noteworthy that Belle, using $253 \, {\rm fb^{-1}}$, has set a limit in the $\mu \nu$ mode
that is within about a factor of three of the SM expectation \cite{belle-lnu}.  Thus, this mode appears likely to
be just beyond the reach of the current $B$-factory experiments.

\begin{table}[h]
\begin{center}
\caption{Upper limits (90\% CL) on ${\cal B}(B^+ \to \ell^+ \nu)$ from Belle\cite{belle-lnu}
and \babarcap\ \cite{babar-lnu},
along with SM expectations. }
\begin{tabular}{|l|c|c|c|}
\hline \textbf{Mode} & \textbf{SM} & \textbf{Belle} &
\textbf{\babarcap}
\\
\hline 
$B^+ \to \mu^+ \nu$ & $\sim 5\times 10^{-7}$ & $<1.6\times 10^{-6}$ & $<6.6\times 10^{-6}$ \\
\hline 
$B^+ \to e^+ \nu$ & $\sim 1\times 10^{-11}$ & $<9.8\times 10^{-7}$ &  \\
\hline
\end{tabular}
\label{table-lnu}
\end{center}
\end{table}

\section{Conclusion}

The rare FCNC decays discussed here are among the most interesting in $B$ physics.  
Considerable experimental progress has been made by  Belle
and \babar.  Yet, a common theme emerges from this discussion of several
different modes.  It is the need for more data.

Significant and probing measurements have become possible in decays involving the
$b \to s \ell^+ \ell^-$ transition, but the results are statistically limited.  
The results reported thus far do not utilize the full data sets of \babar\ or Belle (for
which the ultimate data set will be much larger), but even so, the final results from
\babar\ and even Belle will clearly still be statistics limited.
Other interesting processes such as $B^+ \to \pi^+ \ell^+ \ell^-$ and perhaps
$B^+ \to \mu^+ \nu$ are just beyond the reach of the existing $B$-factories.
To do this physics justice, it seems apparent that a dataset of at least $10 \, {\rm ab^{-1}}$
is needed.  This is one of a number of
strong arguments that make up the physics case for ``super" $B$-factories.

\bigskip

\begin{acknowledgments}
I would like to thank G. Eigen and K. Flood for useful input.
\end{acknowledgments}

\bigskip 

\end{document}